\documentclass[aps,pra,twocolumn,10pt,superscriptaddress]{revtex4-2}
\usepackage[T1]{fontenc}
\usepackage[utf8]{inputenc}
\usepackage[colorlinks=true, citecolor=blue, urlcolor=blue ]{hyperref}
\usepackage{epsfig,newlfont,amssymb,amsfonts,amsmath,bm,subfigure,palatino,mathtools,amsthm,braket,soul,enumitem,color,graphics,graphicx,times,physics,makecell,multirow,float}
\usepackage[normalem]{ulem}
\usepackage{algorithm}
\usepackage{algpseudocode}
\usepackage[euler]{textgreek}
\usepackage{multirow}
\usepackage{braket}

\newcolumntype{P}[1]{>{\centering\arraybackslash}p{#1}}
\graphicspath{{Figures/}{Dynamics/}}

\begin{document}
\title{
Detection and Inhibition of Decoherence Through Collisions: A dynamical approach
}

\author{Sagnik Chakraborty}
\thanks{These authors contributed equally to this work.}
\affiliation{Dipartimento di Ingegneria, Università degli Studi di Palermo, Viale delle Scienze, 90128 Palermo, Italy}

\author{Srijon Ghosh}
\thanks{These authors contributed equally to this work.}
\affiliation{Dipartimento di Ingegneria, Università degli Studi di Palermo, Viale delle Scienze, 90128 Palermo, Italy}
\affiliation{Institute of Fundamental and Frontier Sciences, University of Electronic Science and Technology of China, Chengdu 611731, China}

\author{Federico Amato}
\thanks{These authors contributed equally to this work.}
\affiliation{Dipartimento di Ingegneria, Università degli Studi di Palermo, Viale delle Scienze, 90128 Palermo, Italy}

\author{Gniewomir Sarbicki}
\affiliation{
Institute of Physics, Faculty of Physics,
Astronomy and Informatics,
Nicolaus Copernicus University,
Grudzi\k{a}dzka 5/7, 87--100 Toru\'n, Poland
}

\author{Dariusz Chruściński}
\affiliation{
Institute of Physics, Faculty of Physics,
Astronomy and Informatics,
Nicolaus Copernicus University,
Grudzi\k{a}dzka 5/7, 87--100 Toru\'n, Poland
}

\author{Rosario Lo Franco}
\affiliation{Dipartimento di Ingegneria, Università degli Studi di Palermo, Viale delle Scienze, 90128 Palermo, Italy}

\begin{abstract}

Quantum systems are intrinsically fragile and highly vulnerable to environmental noise, making the preservation of quantum states one of the fundamental challenges in realizing practical quantum technologies. 
We propose two noise-agnostic quantum state protection protocols based on a collisional model. 
The first protocol preserves quantum states of qubits without requiring complete prior knowledge of the underlying noise dynamics, and the second even protects a logical qubit under similar circumstances. These protocols provide robust protection for arbitrary single-qubit and entangled states against general noise processes. 
Furthermore, their capability to safeguard quantum states with an arbitrary number of qubits enhances the reliability, robustness, and scalability of emerging quantum architectures. 
These results establish the proposed framework as a promising approach for practical quantum information processing.

\end{abstract}

\maketitle

\section{Introduction}

The study of open quantum systems has become a cornerstone of contemporary quantum science and technology, since no realistic quantum system can be completely isolated from its surrounding environment. 
Such unavoidable system–environment interactions give rise to dissipation, decoherence, and the leakage of quantum information \cite{breuer2002theory}. 
Among these effects, quantum decoherence remains one of the most significant challenges hindering the realization of practical quantum technologies. 
Preserving quantum states and maintaining their coherence in the presence of environmental disturbances is an exceptionally demanding task. 
Consequently, the suppression of decoherence and the development of efficient quantum error-correction protocols have become essential prerequisites for reliable quantum computation and the implementation of long-lived quantum memories \cite{zaki2013, jing2014}. 
Over the past decades, considerable theoretical and experimental efforts have been dedicated to understanding \cite{darek2010, franco2013, nina2017, LoFranco2017, Aolita_2015} and devising strategies that mitigate environmental effects and safeguard fragile quantum states against decoherence. 
Among the most prominent approaches for combating decoherence are decoherence-free subspaces  \cite{lidar1998, Lidar_2003, Lidar_2014}, dynamical decoupling \cite{viola1999, lofranco2014, yuan2022, morong2023}, and quantum error correction and mitigation \cite{terhal2015, cai2023,li2017, roffe2019, endo2018}. 
Decoherence-free subspaces protect quantum information by exploiting symmetries in system–environment interactions, while dynamical decoupling suppresses decoherence through controlled operations that average out environmental effects. 
Quantum error mitigation, on the other hand, reduces noise-induced errors via characterization and post-processing techniques without the full overhead of quantum error correction. 

It is important to note that although the preservation of quantum coherence is a crucial requirement for maintaining quantum advantages, many quantum information-processing tasks ultimately demand the protection of the complete quantum state. 
Indeed, preserving the complete quantum state \cite{Korotkov2010, Kim2012, wang2014} ensures the retention of both population and coherence information, making quantum state preservation a significant and challenging objective in the presence of environmental noise. 
Moreover, many of the aforementioned approaches rely, either explicitly or implicitly, on prior knowledge of the underlying noisy dynamics. \\

Motivated by these considerations, in this paper, \textit{we propose a state preservation scheme based on a collisional model \cite{CICCARELLO20221} that protects a quantum state against environmental decoherence through an adaptive strategy, even when the exact nature of the underlying noise is unknown.}
Suppose a quantum system is decohering under the action of an arbitrary noise model, about which we do not have complete knowledge. 
Specifically, only limited information regarding the possible noise processes acting on the system is available.
To prevent the effect of the environment, it is first necessary to identify the specific noisy dynamics that are affecting the system, and then apply a suitable correction procedure to mitigate its impact on the quantum system. 
In order to achieve this goal, we use an adaptive protocol using a collisional model (CM) where the ancillas of the CM act in a dual mode of assessment and correction.
In our protocol, a sequence of ancillas interacts sequentially with the system through collisional dynamics. 
Alternate ancillas are measured after their interaction with the system to extract information about the underlying noise. 
Based on the outcome of each measurement, the state of the subsequent ancilla and its interaction with the system are adaptively chosen such that the resulting collision effectively implements the Petz recovery map \cite{Petz1986} on the system. 
This adaptive recovery mechanism counteracts the decoherence induced by the noise, thereby preserving the quantum state.


The paper is organized as follows. In Sec. \ref{protocol}, we introduce the proposed protocol and formulate it as a quantum game. In Sec. \ref{performance}, we analyze its performance and applicability in both single-qubit and multi-qubit settings. Finally, we summarize our findings and present our conclusions in Sec. \ref{conclusion}.

\begin{figure}
    \centering
    \includegraphics[width=\linewidth]{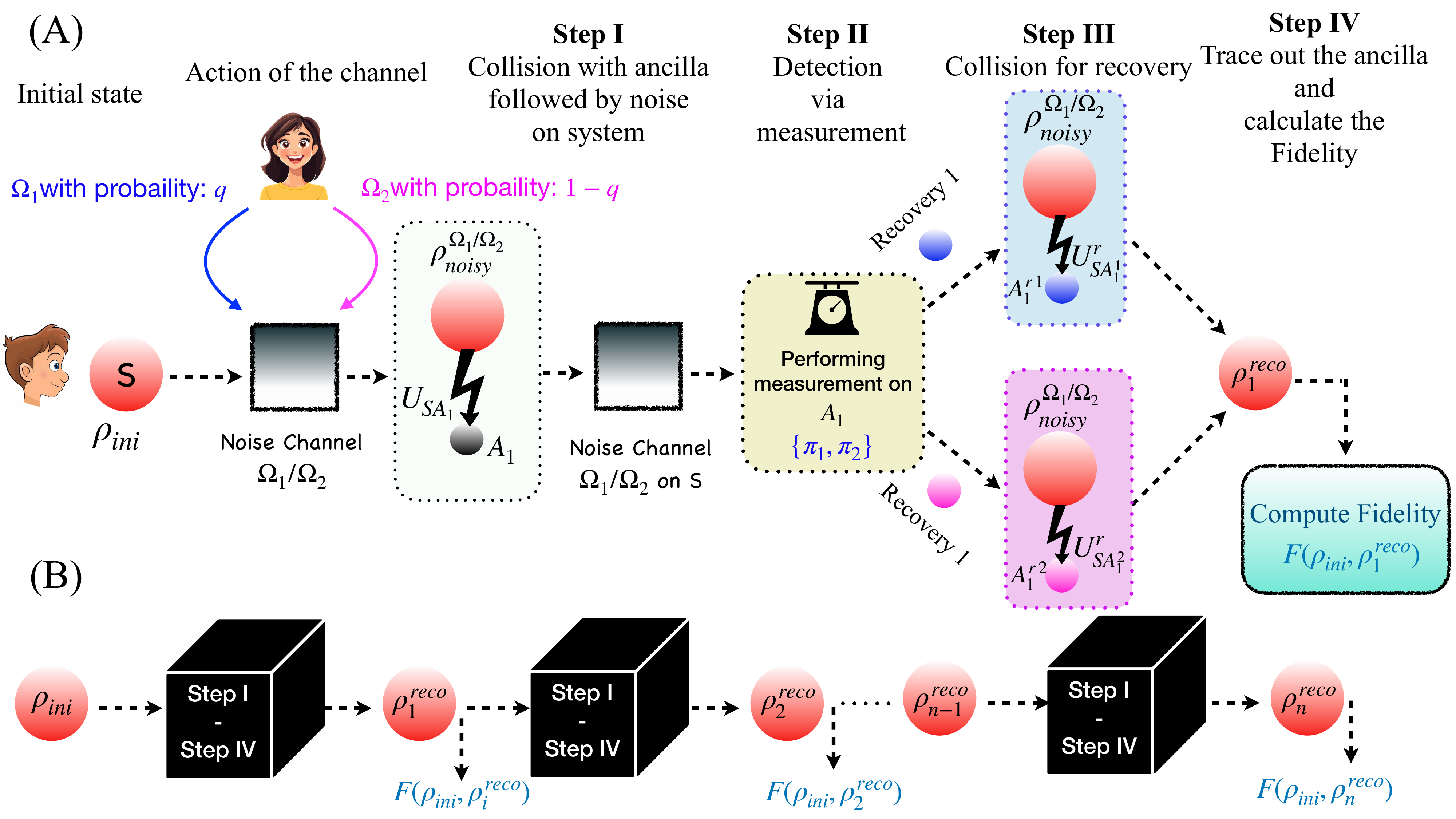}
    \caption{Schematic representation of the proposed protocol. (A) is the illustration of a single run of the protocol, and (B) represents the complete protocol after multiple runs.}
    \label{fig:protocol-scheme}
\end{figure}

\section{Protocol}
\label{protocol}

Our protocol can be best described in the form of a quantum game where Alice chooses a noisy environment $\Omega$ for the system $S$ from a set $\mathcal{O}=\{\Omega_1,\Omega_2\}$ with probability $\mathcal{P}$ = $\{q, 1-q\}$. 
The particular choice made by Alice is unknown to Bob. However, Bob knows about $\mathcal{O}$ and $\mathcal{P}$. 
To start the game, Bob begins with a fixed initial state of the system $S$, denoted by $\rho_\text{ini}$, and allows it to go through a noisy channel $\Omega$, chosen by Alice (either $\Omega_1$ or $\Omega_2$) from  $ \mathcal {O} $ with probability $\mathcal{P}$. 
The resultant state becomes $\rho_\text{noise}=\Omega[\rho_\text{ini}]$. 
Afterwards, Bob uses a CM to assess and preserve the state by eliminating the effect of decoherence caused by $\Omega$. The ancillas of the CM are grouped in pairs. 
In each cycle, the first ancilla assesses the noise and the second corrects it.
If at the end of $(d-1)-$th cycle the system state is $\rho^\text{reco}_{d-1}$, then in the $d-$th cycle, Bob sends an ancilla $A_d$ in state $\eta_d$ (the detection ancilla) to interact with $S$ through the interaction Hamiltonian $H_{SA_d}$. 
Following the interaction, the joint system-ancilla state gets correlated, and Bob again applies the same noise $\Omega$ on the system. The joint resultant state $\rho'_{SA_d}$ reads as
\begin{equation}
    \label{collision_1_on_d}\rho'_{SA_d}=\big(\Omega\otimes\mathbb{I} \big)\big[U_{SA_d}(\rho^\text{reco}_{d-1}\otimes\eta_d)U_{SA_d}^{\dagger}\big].
\end{equation}

Here $U_{SA_d} = \exp (- i H_{SA_d} \Delta t)$, where $\Delta t$ is the time of interaction. Note that the action of $\Omega$ on the system side reduces the correlation between $S$ and $A_d$, and we get a weakly correlated joint state. Having this state, Bob performs a minimum-error discrimination measurement by constructing a POVM $\{\pi_1,\pi_2\}$ on $A_d$. $\{\pi_1,\pi_2\}$ is chosen so that it discriminates with minimum error \cite{Bae_2015} between two ancilla states $\{\eta_d^1, \eta_d^2\}$, occurring with probability $\mathcal{P}$, where $\eta_d^1$ and $\eta_d^2$ are the possible states of the ancilla if the choice of noise made by Alice was $\Omega_1$ and $\Omega_2$, respectively.
That is,
\begin{equation}
    \eta^{1/2}_{d}= \mathrm {Tr}_{S}\Big\{ \big(\Omega_{1/2}\otimes\mathbb{I} \big)\Big[U_{SA_1}\big(\mathcal{N}^{d-1}_{1/2}[\rho_\text{ini}]\otimes\eta_d\big)U_{SA_1}^{\dagger}\Big]\Big\}
\end{equation}
where $\mathcal{N}^{d-1}_{1/2}$ represents the evolution map until the $(d-1)$ cycle if the choice of Alice was $\Omega_1$ or $\Omega_2$, respectively.
This measurement outcome guides Bob to assess 
whether Alice chose $\Omega_1$ or $\Omega_2$, respectively (see Appendix \ref{algorithm}).

Also, due to this measurement performed by Bob, the state of the system ($\mathrm{Tr}_{A_{d}}[\rho'_{SA_d}]$) collapses to the state ($\mathrm{Tr}_{A_{d}}[\beta_{SA_d}]$) due to the correlation (although weak) between the system and the ancilla, where the post-measurement joint state of the system-ancilla is given by
\begin{align}
    \beta_{SA_d} = \frac{(\openone \otimes\pi) \rho'_{SA_d} (\openone\otimes\pi_)}{\text{Tr}\big((\openone\otimes\pi) \rho'_{SA_d} (\openone\otimes\pi_)\big)},
\end{align}
if the measurement outcome corresponds to the POVM element $\pi\in\{\pi_1,\pi_2\}$. 
Depending on the measurement result, Bob chooses the second ancilla ${A^r_d}^{1/2}$ (the recovery ancilla) in state ${\eta^r_d}^{1/2}$ to collide with $S$ through interaction $U^r_{SA_d^{1/2}}$ such that the action of this collision is a Petz recovery map on $S$ and the final state after the recovery becomes $\rho^\text{reco}_{d}$ (see Fig.~\ref{fig:protocol-scheme}). 
The resultant action of the recovery map is given by 
\begin{align}
    \text{Tr}_{{A^r_d}^{1/2}}\big[U^r_{SA_d^{1/2}}(\bullet\otimes{\eta^r_d}^{1/2}){U^r_{SA_d^{1/2}}}^{\dagger}\big]=\Lambda_\text{Petz}(\mathcal{N}'^{d-1}_{1/2},\rho_\text{ini})[\bullet]
\end{align} 
where $\Lambda_\text{Petz}(\mathcal{N},\rho_\text{ini})$ is the Petz recovery map given by $\Lambda_\text{Petz}[\bullet]=\sqrt{\rho_\text{ini}}\,\mathcal{N}^{\dagger}\big[\mathcal{N}[\rho_\text{ini}]^{-1/2}\bullet\mathcal{N}[\rho_\text{ini}]^{-1/2}\big]\sqrt{\rho_\text{ini}}$.
Note that
\begin{align}
\label{map for Petz}
    \mathcal{N}'^{d-1}_{1/2} = \Lambda_\text{cmplt}^{1/2}\circ\Omega_{1/2}\circ\Theta\circ\mathcal{N}^{d-1}_{1/2},
\end{align}
with $\Theta[\bullet]=\text{Tr}_{A_d}\big[U_{SA_d}(\bullet\otimes\eta_d)U_{SA_d}^{\dagger}\big]$ and $\Lambda_\text{cmplt}^{1/2}$ is a quantum channel such that $\Lambda_\text{cmplt}^{1/2}\big[\text{Tr}_{A_d}( \rho'^{1/2}_{SA_d})\big]= \text{Tr}_{A_d}(\beta_{SA_d}^{1/2})$, where $\rho'^{1/2}_{SA_d}$ is obtained by choosing $\Omega_{1/2}$ in place of $\Omega$ in Eq. (\ref{collision_1_on_d}).
A very simple construction of $\Lambda_\text{cmplt}^{1/2}$ for any two states $\{\rho_1,\rho_2\}$ such that $\Lambda_\text{cmplt}^{1/2}[\rho_1]=\rho_2$ is $\Lambda_\text{cmplt}^{1/2}[X]=(\text{Tr}X)\,\rho_2$. 
A non-trivial construction is also given in Appendix \ref{algorithm}.
We explain in detail in Appendix \ref{Stinespring dilation} how any quantum channel can be generated through $U^r_{SA_d}$ and $\eta^r_d$ using Stinespring dilation. 
In short, for a qubit channel $\Lambda[X]=\sum_{i=1}^4K_iXK_i^{\dagger}$ we can choose $\eta_d^r=\ket{00}\bra{00}$ and 
\begin{equation}
    U^r_{SA_d}= \begin{bmatrix}
        K_0 & . & . & . \\
        K_1 & . & . & . \\
        K_2 & . & . & . \\
        K_3 & . & . & . 
    \end{bmatrix}
\end{equation}
where $U^r_{SA_d}$ is the unitary completion of a block matrix with $K_i$'s as the first column. 
This process is repeated over $n$-cycles of alternate measurement-recovery and finally, the fidelity $\mathcal{F} = \text{Tr}(\sqrt{\rho_\text{ini}}\,\rho_n^{reco}\sqrt{\rho_\text{ini}})$ of the final $\rho_{n}^{reco}$ is calculated with respect to $\rho_\text{ini}$.\\

Note here that the most important part of this protocol is how on the $d-$th cycle, Bob chooses $\Lambda_\text{Petz}$ (or $U^r_{SA_d}$ and $\eta^r_d$) based on the measurement outcome. 
We present Bob's strategy in the following way: If, after $d$-cycles of measurement-recovery $c_1$ and $c_2$ represents the number of times outcome ``$1$"  or ``$2$" has occurred, then Bob chooses $\Lambda_\text{Petz}$  corresponding to $\Omega_1$ ($\Omega_2$) in Eq. (\ref{map for Petz}) if $c_1 > c_2 $ ($c_1 < c_2 $). 
If $c_1 = c_2 $, Bob preserves the trend and repeats the choice of the last cycle. 
A detailed algorithm for this protocol is given in Appendix~\ref{algorithm}.

\section{Performance of the protocol}
\label{performance}
In this section, we evaluate the performance of our protocol. We begin by using the CM protocol to detect environmental noise and protect an arbitrary single-qubit state. 
Then we investigate the performance of the proposed protocol in detecting environmental noise and protecting multi-qubit states, notably, entangled states. 
Finally, we discuss how to protect a $2-$dimensional subspace of $N-$qubit states. 
 
\begin{figure}
    \includegraphics[width=0.47\textwidth]{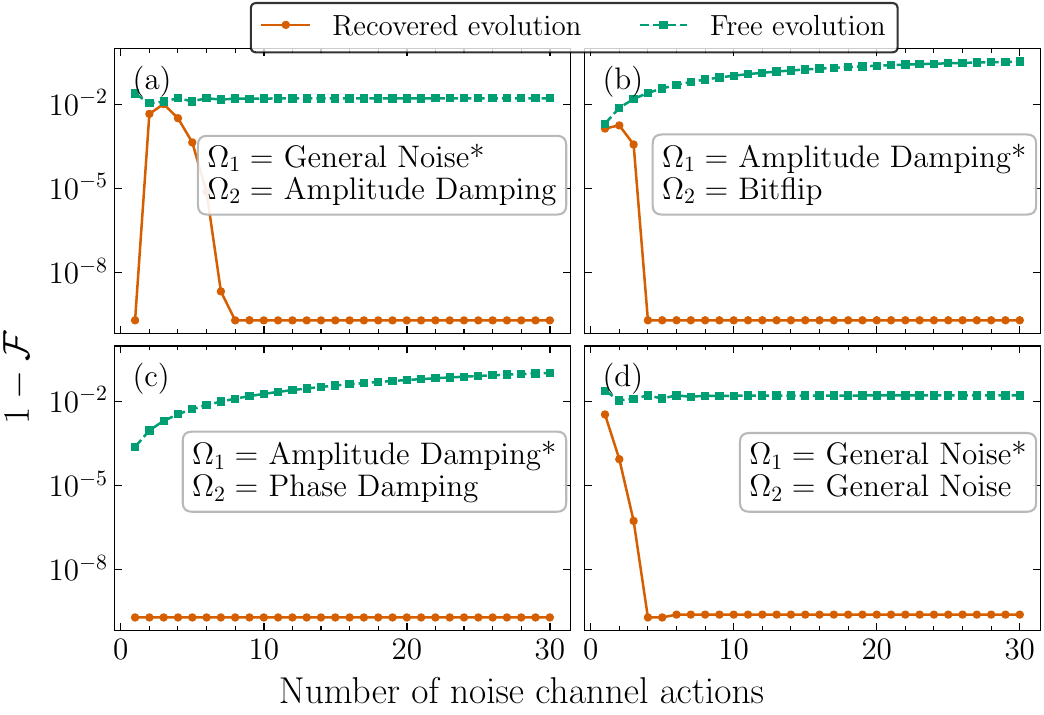}
    \caption{Comparison in terms of infidelity between the actual free evolution of a single-qubit quantum state under noise (marked as $``*"$) and the recovered evolution for different noise pairs. a) Noises are \emph{general noise} and \emph{amplitude damping noise}. Here $\gamma = 1.0$ indicates the strength of the noise, b) $\mathcal{O}$ = \{\emph{Bitflip}, \emph{Amplitude Damping}\}, $\gamma=3.0$, c) $\mathcal{O}$  =  \{\emph{Amplitude Damping}, \emph{Phase Damping}\}, $\gamma=1.0$, and d) Noises are two different instances of \emph{general noise}, Here we take $\Delta t = \pi/100$ and $g = 1$. $\mathcal{P} = (0.5,0.5)$.}
    \label{fig:one-qubit-autorecovery}
\end{figure}

\textit{Recovery of a single qubit:} To assess the effectiveness of our protocol, we first consider the problem of detecting the noise channel acting on a single qubit and subsequently mitigating its effect. 
The initial state of the qubit is taken to be an arbitrary mixed state from the Bloch sphere, given by $\rho_\text{ini} = \frac{1}{2}\left(\mathbb{I} + \mathbf{r}\cdot\boldsymbol{\sigma}\right)$, with $r_x^2 + r_y^2 + r_z^2 \leq 1$. 
This is the quantum state that has to be protected. 
$\boldsymbol{\sigma}=(\sigma_x,\sigma_y,\sigma_z)$ is the Pauli vector and $\mathbf{r}=(r_x,r_y,r_z)$ is the Bloch vector. 
The condition $|\mathbf{r}| \leq 1$ ensures the positivity of the density operator. 
Specifically, we analyze four different configurations of $\mathcal{O}$, with $\mathcal{P} = (0.5,0.5)$ as shown in Fig.~\ref{fig:one-qubit-autorecovery}. Note that our protocol is valid for any values of $q \in [0,1]$. 
This enables us to evaluate the effectiveness of our protocol both in the presence of arbitrary noise processes (see Appendix~\ref{general_kraus} for details) and for computationally relevant noise models of practical interest (for e.g., Amplitude-damping channel, Dephasing channel, and bit-flip channel). 
In the quantum game, $``\ast"$ denotes the actual noisy channel chosen by Alice from the set $\mathcal{O}$, however Bob only knows $\mathcal{O}$ and $\mathcal{P}$. The free evolution indicates the evolution of the system without the recovery protocol. 
This serves as a benchmark of our protocol. 
In the $d$-th cycle of the protocol, the interaction between the system (S) and the detection ancilla ($A_{d}$) is taken to be $H_\text{int} = g(\sigma^{+}_{S} \sigma^{-}_{A_{d}} + \sigma^{-}_{S} \sigma^{+}_{A_{d}})$. 
As illustrated in Fig.~\ref{fig:one-qubit-autorecovery}, when the protocol is not applied (free evolution), the quantum state progressively decoheres and departs from the initial state $\rho_{\mathrm{ini}}$ and ultimately converges to a noise-induced steady state. 
This is obvious as no recovery protocol is applied to the system and infidelity, ($1-\mathcal{F}$) becomes $0.1$. 
By contrast, the proposed detection-and-recovery protocol enables the preservation of the initial state over long timescales. 
Remarkably, this perfect recovery can be achieved without complete prior knowledge of the particular noise channel acting on the system. Although for certain combinations of noise models (see Fig.~\ref{fig:one-qubit-autorecovery}), the infidelity increases, indicating the evolved state goes away from the target state, showing a bump in the infidelity. 
Since the protocol is probabilistic by nature as Bob applies the recovery depending on the measurement outcome, it may happen that wrong recovery is being applied to the state that takes it away from the original state. 
However, the protocol is so designed that successive applications of the protocol at finite intervals repeatedly restore the system to $\rho_\text{ini}$ and infidelity becomes $10^{-9}$. 
Therefore, the detrimental effects of noise can be effectively suppressed for arbitrarily long durations. 
Note that the construction of $\Lambda_\text{cmplt}^{1/2}$ is explained in Appendix \ref{algorithm}. 

\begin{figure}
    \includegraphics[width=0.47\textwidth]{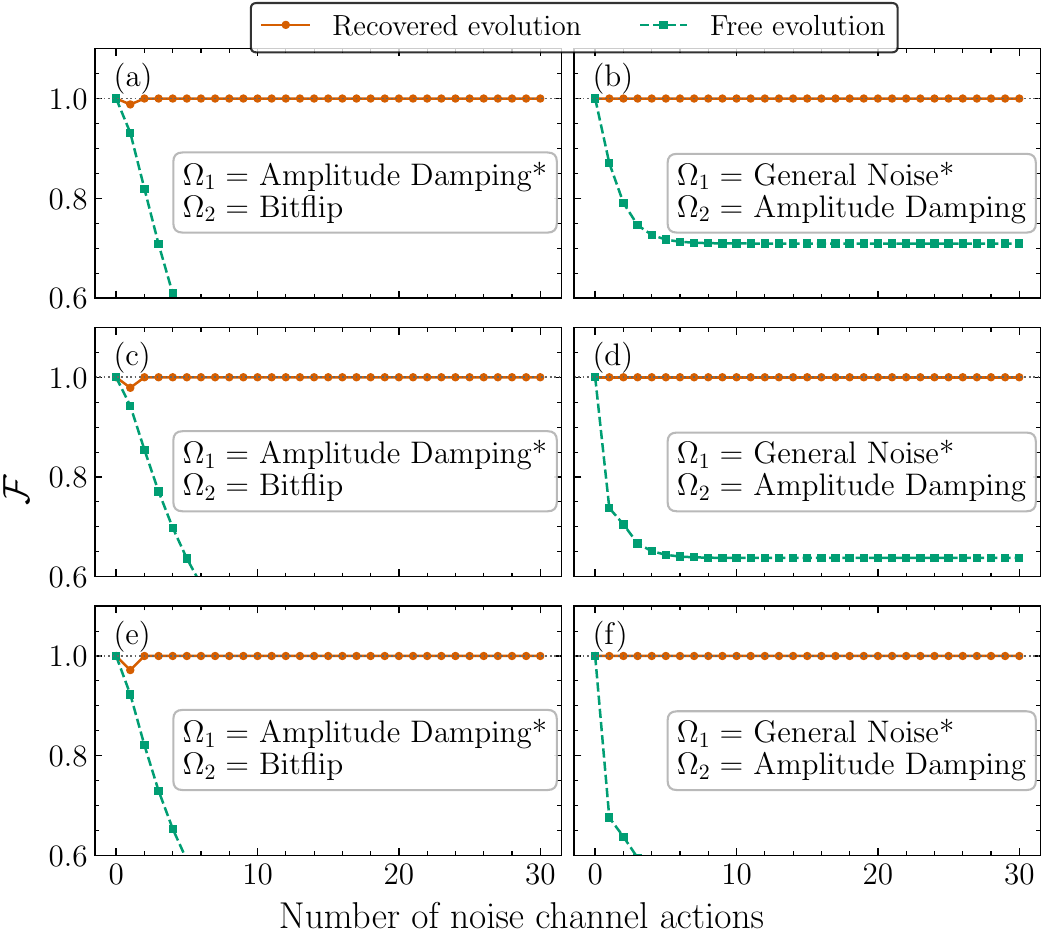}
    \caption{Recovery of an $N$-qubit entangled state in the presence of unknown noise for $\gamma=10.0$. Panels (a) and (b) correspond to the two-qubit case ($N=2$), panels (c) and (d) to the three-qubit case ($N=3$), and panels (e) and (f) to the four-qubit case ($N=4$). All other parameters and simulation details are identical to those used in Fig.~\ref{fig:one-qubit-autorecovery}.}
    \label{fig:multi-qubit-autorecovery}
\end{figure}






\textit{Protection of an entangled state:} 
Let us now consider the recovery of a multi-qubit system. 
Here we consider an $N$-qubit generalized Werner state as the initial state. 
For an $N$-qubit system, consider a parametrized state of the form 
\begin{equation*}
    \rho_{W}=\zeta|GHZ_{N}\rangle \langle GHZ_{N}|+(1-\zeta){\mathbb{I}_{2^{N}}}/{2^{N}},  
\end{equation*}
where $\zeta \in [0,1]$ quantifies the relative contribution of the GHZ state, 
\begin{equation*}
    |GHZ_{N}\rangle = \frac{|0\rangle^{\otimes N} + |1\rangle^{\otimes N}}{\sqrt{2}}.
\end{equation*}
The entanglement properties of $\rho_W$ depend on the mixing parameter $\zeta$. 
In particular, the state is fully separable below a critical threshold and entangled above it, according to
\begin{equation}
    \begin{cases}
    \zeta > \frac{1}{1+2^{N-1}} \Rightarrow \text{entangled}\\
    \zeta \leq \frac{1}{1+2^{N-1}} \Rightarrow \text{fully separable}
    \end{cases}
\end{equation}

Fig. \ref{fig:multi-qubit-autorecovery} shows the recovery of an entangled state in the presence of noise via our protocol.
Here we consider the detection interaction between $A_{d}$ and $S$ to be $H_{int} = g(\mathcal{Q}^{+}_{S} \otimes \sigma^{-}_{A_{d}} + \mathcal{Q}^{-}_{S} \otimes \sigma^{+}_{A_{d}})$, where $\mathcal{Q}_{\pm} = (S_{x} \pm iS_{y})/2$ with $S_{x}$, $S_{y}$ being the higher-dimensional Pauli operators. 
Remarkably, the protocol can protect the mixed entangled state of an arbitrary number of qubits with fidelity unity even against any pair of $\mathcal{O}$.

\subsection*{Protecting a logical subspace}
Next, we extend our protocol to protect an entire subspace of states~\cite{kwon2022}, namely a logical qubit or a codespace \cite{roffe2019} $\mathcal{C}:=\text{span}\{\ket{0}^{\otimes N},\ket{1}^{\otimes N}\}$. 
We consider $N$-qubit code states in the following form
\begin{equation}
\ket{\psi_{\mathcal{C}}^N}
=
\alpha_0 \ket{0}^{\otimes N}
+
\alpha_1 e^{-i\phi}\ket{1}^{\otimes N}
\end{equation}
where $|\alpha_0|^2+|\alpha_1|^2=1$. The noise acting on the entire $N-$qubit system is given by $\Lambda_X^N=e^{\tau \mathcal{L}_X}\otimes..\otimes e^{\tau \mathcal{L}_X}$ where
\begin{equation}
    \label{N-qubit_noise}\mathcal{L}_X(\rho)=\gamma_x\mathcal{D}(\sigma_x)[\rho]
\end{equation}
with $\mathcal{D}(Z)[\bullet]=Z\bullet Z^{\dagger}-\frac{1}{2}\{Z^{\dagger}Z, \bullet\}$ for various values of $\gamma_x$. It is important to note here that in our strategy of protecting initial states from the logical subspace, the recovery protocol operated by Bob should be the same irrespective of the initial states. 
We show here that in spite of this, states from the logical subspace are protected with considerably high fidelity. 
In order to explain this in terms of the quantum game, we invoke here a new character---Charlie---whose job is to supply Bob with the initial states and inform Bob which initial state has been supplied. 
Also, at the end of the protocol, Bob hand over final the states to Charlie, and Charlie measures the fidelity of the final state with the initial state that he had supplied to Bob. 
In our protocol, Charlie is dishonest with Bob. 
He supplies different initial states from the code space but tells  Bob that the initial state is always the target state. 
Thus, Charlie selects $\rho_\text{ini}=\ket{\psi^N_{\mathcal{C}}}\bra{\psi^N_{\mathcal{C}}}$, where $\ket{\psi_{C}^N}\in\mathcal{C}$, and tells Bob that the initial state is $\sigma=e^{-\beta H}/Z$ with $H=-\sum_{j>i}\sigma_z^{(i)}\sigma_z^{(j)}$ and $Z$ the partition function. 
As a result, Bob runs the protocol according to the initial state $\sigma$. 

We also slightly modify the protocol for this case. 
The presence of a strong measurement tends to collapse the state to a single final state in each recovery cycle, thereby destroying the average recovery fidelity for any $\rho_\text{ini}$ from $\mathcal{C}$. 
As a result, we make the measurement unsharp and make the interaction during the first collision of each cycle very weak, i.e. we modify the POVM elements to $\pi_i^\text{weak}=w \openone/2+(1-w) \pi_i$ where $0\le w \le 1$. We also modify the collision unitary to $U^\text{weak}_{SA_d}\approx \openone$ by keeping the collision time $\Delta t$ small. 

\begin{figure}
    \centering
    \includegraphics[width=0.8\linewidth]{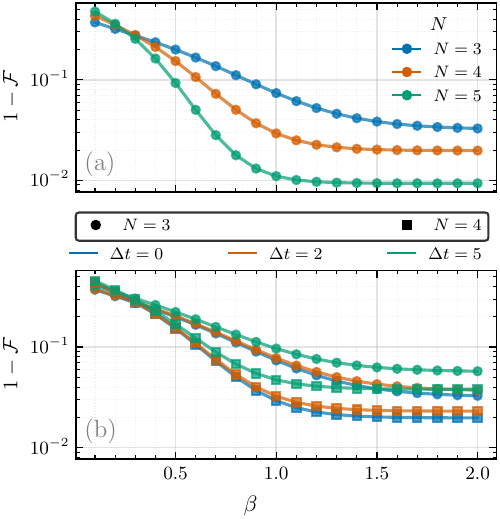}
    \caption{Average infidelity $1-\mathcal{F}$ with respect to the state parameter $\beta$ of the first recovery collision over a sample of 100 Haar-uniformly generated 4-qubit code space states ($\in\mathcal{C}$). Here, two noise channels are bit-flip channels with $\gamma_x^1=\gamma_x^2=1.0$. (a) Illustrates infidelity for different values of $N$ and (b) shows the variation with $\Delta t$. Here $w = 0.9$. }

    \label{fig:same-noises}
\end{figure}

We examine two scenarios: $(i)$ Protecting a logical subspace with single noise, i.e. $\gamma_x^{1}=\gamma_x^{2}$ in Eq. (\ref{N-qubit_noise}), and $(ii)$ protecting a logical subspace when the nature of the noises are similar but the strengths are different, i,e. $\gamma_x^{1}\neq \gamma_x^{2}$
in Eq. (\ref{N-qubit_noise}). 
In both cases, we analyze the recovery infidelity $1 - \mathcal{F}$ averaged over an ensemble of $100$ Haar-uniformly generated states from $\mathcal{C}$ for one cycle of recovery, i.e. $n=1$. 
In the first scenario, for $N = 5$, the average infidelity becomes $10^{-2}$ indicating that the state is almost fully recovered in a single cycle (see Fig.\ref{fig:same-noises} (a)). 
Although we see that increasing the number of qubits provides better fidelity of the recovered state with respect to the initial state, increasing the interaction time $\Delta t$ has a detrimental effect for a fixed value of $N$ (see Fig. \ref{fig:same-noises}(b)).
The high recovery fidelity $\mathcal{F}$ for the logical subspace implies that this technique is very useful for a single known noise. 

\begin{figure}
    \centering
    \includegraphics[width=0.8\linewidth]{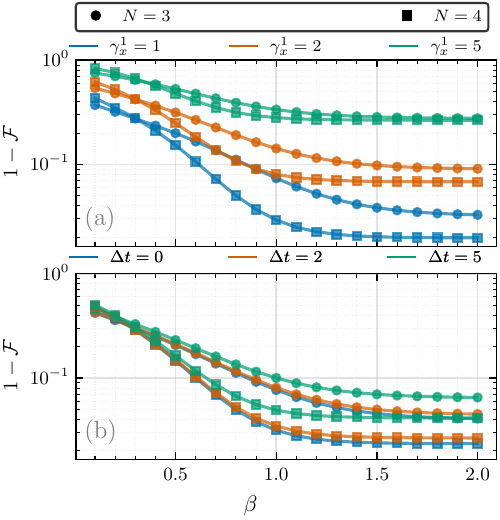}
    \caption{Trends of infidelity with $\beta$. Unlike Fig. \ref{fig:same-noises}, here $\gamma_{x}^{1} \ne \gamma_{x}^{2}$. In (a), $\gamma_{x}^{2}$ is fixed at unity, and variation of $1-\mathcal{F}$ is presented for different values of $\gamma_{x}^{1}$, and the number of qubits $N$, for $\Delta t = 0.1$. (b) Illustrates the same for different values of $\Delta t$ with $\gamma_{x}^{1} = 1$ and $\gamma_{x}^{2} = 2$.}
    \label{fig:splitting-noises}
\end{figure}

\begin{figure}
    \centering
    \includegraphics[width=0.75\linewidth]{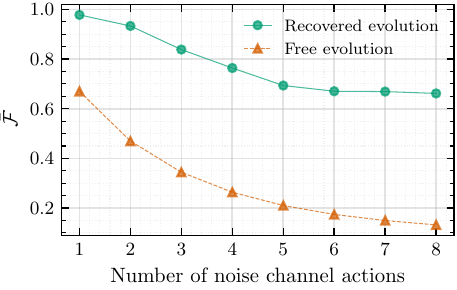}
    \caption{Average fidelity $\overline {\mathcal{F}}$ as a function of the number of noise channel actions. The average is computed over 100 Haar-random states drawn from the code space $\mathcal{C}$. The recovered evolution is compared with the corresponding free evolution, in which the same states are subject to the noise channel without recovery. The applied noise is a bitflip channel with $\gamma_x^1=1.0$, and the recovery protocol starts with equal prior probabilities to the two candidate noise strengths, $\gamma_x^1=1.0$ and $\gamma_x^2=2.0$. The protocol uses a thermal reference state $\sigma$ with inverse temperature $\beta=2.0$. Collision time is small and measurements are weak, $\Delta t=0.1$ and $w=0.9$.}
    \label{fig:codespace-recovery}
\end{figure}

Scenario $(ii)$: Here we slowly start to vary the two noises and check how low the infidelity $1-\mathcal{F}$ falls depending on the difference in values of $\gamma_x^{1}$ and $\gamma_x^{2}$. 
From the plot Fig.~\ref{fig:splitting-noises}(a) and~\ref{fig:splitting-noises}(b), we see that the infidelity falls more rapidly when $\gamma_x^2$ is closer to $\gamma_x^1$ and $\Delta t$ is small. 
This suggests that a weak interaction during the first collision in each cycle is most useful. 
In Fig.~\ref{fig:codespace-recovery}, the recovery protocol for the scenario \emph{(ii)} is applied iteratively over several noise-recovery cycles. 
Since the noise channel acts again on the recovered system at each cycle, a gradual decrease in fidelity is expected. 
Nevertheless, the protocol maintains an average fidelity above 80\% during the first few cycles and subsequently approaches a plateau around 65\%. 
By contrast, in the absence of recovery, the fidelity decreases rapidly with repeated applications of the noise channel, indicating a substantial loss of the information encoded in the initial states.

\section{Conclusion}
\label{conclusion}

In this work, we have introduced an adaptive quantum state protection framework based on collisional dynamics that enables the identification and mitigation of environmental noise without requiring complete prior knowledge of the underlying noisy process. The proposed protocol exploits the dual role of ancillary systems, which are sequentially employed to extract information about the noise and implement a suitable recovery operation through the Petz recovery map.

We have demonstrated the effectiveness of the protocol for arbitrary single-qubit states subjected to different noise channels, achieving near-perfect recovery even after repeated exposure to environmental noise. Furthermore, we have extended the framework to multipartite systems and shown that it can successfully protect entangled states against decoherence. Going beyond the preservation of individual quantum states, we have also investigated the protection of an entire logical subspace. By introducing weak measurements and sufficiently short system–ancilla interactions, we have shown that quantum information encoded within a logical qubit can be preserved with high fidelity, even when the strength of the underlying noise is not precisely known.

Our findings establish collisional dynamics as a promising framework for adaptive quantum state protection, bridging environmental noise identification and quantum recovery within a unified protocol. The ability to mitigate decoherence with incomplete information about the environment, together with its applicability to multipartite states and encoded quantum information, opens up potential avenues for developing noise-resilient quantum information processing and quantum memory architectures.

\begin{acknowledgments}
The authors acknowledge support by MUR (Ministero dell’Universit\`{a} e della Ricerca) through the PNRR Project ICON-Q – Partenariato Esteso NQSTI – PE00000023 – Spoke 2 – CUP: J13C22000680006. R.L.F. also acknowledges support by MUR through the PNRR Project QUANTIP – Partenariato Esteso NQSTI – PE00000023 – Spoke 9 – CUP: E63C22002180006. 
\end{acknowledgments}

\bibliography{bibfile.bib}

@article{lidar1998,
  title = {Decoherence-Free Subspaces for Quantum Computation},
  author = {Lidar, D. A. and Chuang, I. L. and Whaley, K. B.},
  journal = {Phys. Rev. Lett.},
  volume = {81},
  issue = {12},
  pages = {2594--2597},
  numpages = {0},
  year = {1998},
  month = {Sep},
  publisher = {American Physical Society},
  doi = {10.1103/PhysRevLett.81.2594},
  url = {https://link.aps.org/doi/10.1103/PhysRevLett.81.2594}
}

@inbook{Lidar_2003,
   title={Decoherence-Free Subspaces and Subsystems},
   ISBN={9783540448747},
   ISSN={0075-8450},
   url={http://dx.doi.org/10.1007/3-540-44874-8_5},
   DOI={10.1007/3-540-44874-8_5},
   booktitle={Irreversible Quantum Dynamics},
   publisher={Springer Berlin Heidelberg},
   author={Lidar, Daniel A. and Birgitta Whaley, K.},
   year={2003},
   pages={83–120} }

@book{lidar_2014,
   title={Quantum Information and Computation for Chemistry},
   ISBN={9781118742631},
   ISSN={1934-4791},
   url={http://dx.doi.org/10.1002/9781118742631},
   DOI={10.1002/9781118742631},
   journal={Advances in Chemical Physics},
   publisher={Wiley},
   year={2014},
   month=Feb }

@article{yuan2022,
  title = {Preserving multilevel quantum coherence by dynamical decoupling},
  author = {Yuan, Xinxing and Li, Yue and Zhang, Mengxiang and Liu, Chang and Zhu, Mingdong and Qin, Xi and Vitanov, Nikolay V. and Lin, Yiheng and Du, Jiangfeng},
  journal = {Phys. Rev. A},
  volume = {106},
  issue = {2},
  pages = {022412},
  numpages = {7},
  year = {2022},
  month = {Aug},
  publisher = {American Physical Society},
  doi = {10.1103/PhysRevA.106.022412},
  url = {https://link.aps.org/doi/10.1103/PhysRevA.106.022412}
}

@article{viola1999,
  title = {Dynamical Decoupling of Open Quantum Systems},
  author = {Viola, Lorenza and Knill, Emanuel and Lloyd, Seth},
  journal = {Phys. Rev. Lett.},
  volume = {82},
  issue = {12},
  pages = {2417--2421},
  numpages = {0},
  year = {1999},
  month = {Mar},
  publisher = {American Physical Society},
  doi = {10.1103/PhysRevLett.82.2417},
  url = {https://link.aps.org/doi/10.1103/PhysRevLett.82.2417}
}

@book{breuer2002theory,
  title={The theory of open quantum systems},
  author={Breuer, Heinz-Peter and Petruccione, Francesco},
  year={2002},
  publisher={OUP Oxford}
}

@article{wang2014,
  title = {Protecting quantum states from decoherence of finite temperature using weak measurement},
  author = {Wang, Shu-Chao and Yu, Zong-Wen and Zou, Wen-Jie and Wang, Xiang-Bin},
  journal = {Phys. Rev. A},
  volume = {89},
  issue = {2},
  pages = {022318},
  numpages = {8},
  year = {2014},
  month = {Feb},
  publisher = {American Physical Society},
  doi = {10.1103/PhysRevA.89.022318},
  url = {https://link.aps.org/doi/10.1103/PhysRevA.89.022318}
}

@article{Korotkov2010,
  title = {Decoherence suppression by quantum measurement reversal},
  author = {Korotkov, Alexander N. and Keane, Kyle},
  journal = {Phys. Rev. A},
  volume = {81},
  issue = {4},
  pages = {040103(R)},
  numpages = {4},
  year = {2010},
  month = {Apr},
  publisher = {American Physical Society},
  doi = {10.1103/PhysRevA.81.040103},
  url = {https://link.aps.org/doi/10.1103/PhysRevA.81.040103}
}

@article{Kim2012,
  author  = {Kim, Yong-Su and Lee, Jong-Chan and Kwon, Osung and Kim, Yoon-Ho},
  title   = {Protecting entanglement from decoherence using weak measurement and quantum measurement reversal},
  journal = {Nature Physics},
  volume  = {8},
  number  = {2},
  pages   = {117--120},
  year    = {2012},
  doi     = {10.1038/nphys2176}
}

@article{CICCARELLO20221,
title = {Quantum collision models: Open system dynamics from repeated interactions},
journal = {Phys. Rep.},
volume = {954},
pages = {1-70},
year = {2022},
note = {Quantum collision models: Open system dynamics from repeated interactions},
issn = {0370-1573},
doi = {https://doi.org/10.1016/j.physrep.2022.01.001},
url = {https://www.sciencedirect.com/science/article/pii/S0370157322000035},
author = {Francesco Ciccarello and Salvatore Lorenzo and Vittorio Giovannetti and G. Massimo Palma}
}

@article{kwon2022,
  title = {Reversing Lindblad Dynamics via Continuous Petz Recovery Map},
  author = {Kwon, Hyukjoon and Mukherjee, Rick and Kim, M. S.},
  journal = {Phys. Rev. Lett.},
  volume = {128},
  issue = {2},
  pages = {020403},
  numpages = {7},
  year = {2022},
  month = {Jan},
  publisher = {American Physical Society},
  doi = {10.1103/PhysRevLett.128.020403},
  url = {https://link.aps.org/doi/10.1103/PhysRevLett.128.020403}
}

@article{Petz1986,
  author   = {Petz, D{\'{e}}nes},
  title    = {Sufficient subalgebras and the relative entropy of states of a von Neumann algebra},
  journal  = {Communications in Mathematical Physics},
  year     = {1986},
  volume   = {105},
  number   = {1},
  pages    = {123--131},
  month    = {Mar},
  doi      = {10.1007/BF01212345},
  url      = {https://doi.org/10.1007/BF01212345},
  issn     = {1432-0916}
}

@article{terhal2015,
  title = {Quantum error correction for quantum memories},
  author = {Terhal, Barbara M.},
  journal = {Rev. Mod. Phys.},
  volume = {87},
  issue = {2},
  pages = {307--346},
  numpages = {40},
  year = {2015},
  month = {Apr},
  publisher = {American Physical Society},
  doi = {10.1103/RevModPhys.87.307},
  url = {https://link.aps.org/doi/10.1103/RevModPhys.87.307}
}

@article{cai2023,
  title = {Quantum error mitigation},
  author = {Cai, Zhenyu and Babbush, Ryan and Benjamin, Simon C. and Endo, Suguru and Huggins, William J. and Li, Ying and McClean, Jarrod R. and O'Brien, Thomas E.},
  journal = {Rev. Mod. Phys.},
  volume = {95},
  issue = {4},
  pages = {045005},
  numpages = {37},
  year = {2023},
  month = {Dec},
  publisher = {American Physical Society},
  doi = {10.1103/RevModPhys.95.045005},
  url = {https://link.aps.org/doi/10.1103/RevModPhys.95.045005}
}

@article{zaki2013,
  title = {Hardware-Efficient Autonomous Quantum Memory Protection},
  author = {Leghtas, Zaki and Kirchmair, Gerhard and Vlastakis, Brian and Schoelkopf, Robert J. and Devoret, Michel H. and Mirrahimi, Mazyar},
  journal = {Phys. Rev. Lett.},
  volume = {111},
  issue = {12},
  pages = {120501},
  numpages = {5},
  year = {2013},
  month = {Sep},
  publisher = {American Physical Society},
  doi = {10.1103/PhysRevLett.111.120501},
  url = {https://link.aps.org/doi/10.1103/PhysRevLett.111.120501}
}

@misc{jing2014,
      title={Overview of Quantum Memory Protection and Adiabaticity Induction by Fast-Signal Control}, 
      author={Jun Jing and Lian-Ao Wu},
      year={2014},
      eprint={1412.1783},
      archivePrefix={arXiv},
      primaryClass={quant-ph},
      url={https://arxiv.org/abs/1412.1783}, 
}

@article{Bae_2015,
doi = {10.1088/1751-8113/48/8/083001},
url = {https://doi.org/10.1088/1751-8113/48/8/083001},
year = {2015},
month = {jan},
publisher = {IOP Publishing},
volume = {48},
number = {8},
pages = {083001},
author = {Bae, Joonwoo and Kwek, Leong-Chuan},
title = {Quantum state discrimination and its applications},
journal = {Journal of Physics A: Mathematical and Theoretical}
}

@article{lofranco2014,
  title = {Preserving entanglement and nonlocality in solid-state qubits by dynamical decoupling},
  author = {Lo Franco, R. and D'Arrigo, A. and Falci, G. and Compagno, G. and Paladino, E.},
  journal = {Phys. Rev. B},
  volume = {90},
  issue = {5},
  pages = {054304},
  numpages = {11},
  year = {2014},
  month = {Aug},
  publisher = {American Physical Society},
  doi = {10.1103/PhysRevB.90.054304},
  url = {https://link.aps.org/doi/10.1103/PhysRevB.90.054304}
}

@article{darek2010,
  title = {Long-time memory in non-Markovian evolutions},
  author = {Chru\ifmmode \acute{s}\else \'{s}\fi{}ci\ifmmode \acute{n}\else \'{n}\fi{}ski, Dariusz and Kossakowski, Andrzej and Pascazio, Saverio},
  journal = {Phys. Rev. A},
  volume = {81},
  issue = {3},
  pages = {032101},
  numpages = {6},
  year = {2010},
  month = {Mar},
  publisher = {American Physical Society},
  doi = {10.1103/PhysRevA.81.032101},
  url = {https://link.aps.org/doi/10.1103/PhysRevA.81.032101}
}

@article{nina2017,
  title={Eternal non-Markovianity: from random unitary to Markov chain realisations},
  author={Megier, Nina and Chru{\'s}ci{\'n}ski, Dariusz and Piilo, Jyrki and Strunz, Walter T},
  journal={Scientific reports},
  volume={7},
  number={1},
  pages={6379},
  year={2017},
  publisher={Nature Publishing Group UK London},
  url ={https://www.nature.com/articles/s41598-017-06059-5#Sec4}
}

@article{morong2023,
  title = {Engineering Dynamically Decoupled Quantum Simulations with Trapped Ions},
  author = {Morong, W. and Collins, K.S. and De, A. and Stavropoulos, E. and You, T. and Monroe, C.},
  journal = {PRX Quantum},
  volume = {4},
  issue = {1},
  pages = {010334},
  numpages = {16},
  year = {2023},
  month = {Mar},
  publisher = {American Physical Society},
  doi = {10.1103/PRXQuantum.4.010334},
  url = {https://link.aps.org/doi/10.1103/PRXQuantum.4.010334}
}

@Inbook{LoFranco2017,
author="Lo Franco, Rosario
and Compagno, Giuseppe",
editor="Fanchini, Felipe Fernandes
and Soares Pinto, Diogo de Oliveira
and Adesso, Gerardo",
title="Overview on the Phenomenon of Two-Qubit Entanglement Revivals in Classical Environments",
bookTitle="Lectures on General Quantum Correlations and their Applications",
year="2017",
publisher="Springer International Publishing",
address="Cham",
pages="367--391",
isbn="978-3-319-53412-1",
doi="10.1007/978-3-319-53412-1_17",
url="https://doi.org/10.1007/978-3-319-53412-1_17"
}

@article{roffe2019,
  title={Quantum error correction: an introductory guide},
  author={Roffe, Joschka},
  journal={Contemporary Physics},
  volume={60},
  number={3},
  pages={226--245},
  year={2019},
  publisher={Taylor \& Francis},
  url ={https://www.tandfonline.com/doi/full/10.1080/00107514.2019.1667078}
}

@article{franco2013,
author = {FRANCO, ROSARIO LO and BELLOMO, BRUNO and MANISCALCO, SABRINA and COMPAGNO, GIUSEPPE},
title = {DYNAMICS OF QUANTUM CORRELATIONS IN TWO-QUBIT SYSTEMS WITHIN NON-MARKOVIAN ENVIRONMENTS},
journal = {International Journal of Modern Physics B},
volume = {27},
number = {01n03},
pages = {1345053},
year = {2013},
doi = {10.1142/S0217979213450537},

URL = { 
    
        https://doi.org/10.1142/S0217979213450537
    
    

},
eprint = { 
    
        https://doi.org/10.1142/S0217979213450537
    
    

}
}

@article{li2017,
  title = {Efficient Variational Quantum Simulator Incorporating Active Error Minimization},
  author = {Li, Ying and Benjamin, Simon C.},
  journal = {Phys. Rev. X},
  volume = {7},
  issue = {2},
  pages = {021050},
  numpages = {14},
  year = {2017},
  month = {Jun},
  publisher = {American Physical Society},
  doi = {10.1103/PhysRevX.7.021050},
  url = {https://link.aps.org/doi/10.1103/PhysRevX.7.021050}
}

@article{endo2018,
  title = {Practical Quantum Error Mitigation for Near-Future Applications},
  author = {Endo, Suguru and Benjamin, Simon C. and Li, Ying},
  journal = {Phys. Rev. X},
  volume = {8},
  issue = {3},
  pages = {031027},
  numpages = {21},
  year = {2018},
  month = {Jul},
  publisher = {American Physical Society},
  doi = {10.1103/PhysRevX.8.031027},
  url = {https://link.aps.org/doi/10.1103/PhysRevX.8.031027}
}

@article{Aolita_2015,
doi = {10.1088/0034-4885/78/4/042001},
url = {https://doi.org/10.1088/0034-4885/78/4/042001},
year = {2015},
month = {mar},
publisher = {IOP Publishing},
volume = {78},
number = {4},
pages = {042001},
author = {Aolita, Leandro and de Melo, Fernando and Davidovich, Luiz},
title = {Open-system dynamics of entanglement:a key issues review},
journal = {Reports on Progress in Physics}
}
\appendix


\section{Recovery Protocol Algorithm}\label{algorithm}
 {\it Algorithm:} Let $\{q_1,q_2\}$ be the two probabilities with which two noise models $\Omega_1,\Omega_2$ for time $\tau$ are sampled. Let us begin by writing the following functions: 
\begin{itemize}
    \item $(\Pi_1,\Pi_2)=$ discrim$(\rho_{test},\rho_1,\rho_2)$. This function returns a POVM $\{\Pi_1,\Pi_2\}$ for optimal discrimination of ensemble $\{\{q_1,\rho_1\},\{q_2,\rho_2\}\}$.
    
    \item $\Lambda_\text{Petz}=$ Petz$(\sigma,\mathcal{N})$. This function returns the Petz recovery map $\Lambda_\text{Petz}[\bullet]=\sigma^{1/2}\mathcal{N}^{\dagger}\big[\mathcal{N}(\sigma)^{-1/2}\,(\bullet)\,\mathcal{N}(\sigma)^{-1/2}\big]\sigma^{1/2}$.
    
    \item $(\eta,U_{SA})=$ collision$(\Lambda)$. This function prepares the collisional model for the map i.e. $\Lambda[\bullet]=\text{Tr}_A\big[U_{SA}\,\big((\bullet) \,\otimes \eta\big)\,U_{SA}^{\dagger}\big]$
    
    \item $count=$ rand$(p_1,p_2)$. This function returns $1$ or $2$ with probability $p_1$ and $p_2$, respectively.

    \item $\Lambda_{cmplt} =$ map-complete $(\alpha,\beta)$. This function returns a CPTP map $\Lambda_{cmplt}$ taking qubit states $\alpha$ to $\beta$.\\
    {\it Construction:} If $\alpha = p_1\ket{\psi_1}\bra{\psi_1} + p_2\ket{\psi_2}\bra{\psi_2}$, and, $\beta = q_1\ket{\phi_1}\bra{\phi_1} + q_2\ket{\phi_2}\bra{\phi_2}$, we have $\Lambda_{cmplt}=\mathcal{U}_1\circ\Gamma\circ\mathcal{U}_2$, where $\mathcal{U}_i[X]=U_i\,X\,U_i^{\dagger}$for $i=1,2$ with $U_1=\ket{\phi_1}\bra{0}+\ket{\phi_2}\bra{1}$, $U_2=\ket{0}\bra{\psi_1}+\ket{1}\bra{\psi_2}$. Thus, need $\Gamma[X]=K_0XK_0^{\dagger}+K_1XK_1^{\dagger}$ s.t. $\Gamma:\begin{bmatrix}
        p_1 & 0 \\
        0 & p_2
    \end{bmatrix}\rightarrow\begin{bmatrix}
        q_1 & 0 \\
        0 & q_2
    \end{bmatrix}$.\\
    $(a)$ If $q_1=p_1$, we choose $\Gamma[X]=X$.\\
    $(b)$ If $q_1<p_1$, we choose $K_0=\begin{bmatrix}
        \sqrt{q_1/p_1} & 0 \\
        0 & 1
    \end{bmatrix}$, $K_1=\begin{bmatrix}
        0 & 0 \\
        \sqrt{1-q_1/p_1} & 0
    \end{bmatrix}$.\\
    $(c)$ If $q_2<p_2$, we choose $K_0=\begin{bmatrix}
        1 & 0 \\
        0 & \sqrt{q_2/p_2}
    \end{bmatrix}$, $K_1=\begin{bmatrix}
        0 & \sqrt{1-q_2/p_2} \\
        0 & 0
    \end{bmatrix}$. 
    So, given an arbitrary dimension $d$, this reads:
    $\beta = \Lambda_\text{cmplt}[\alpha]=\mathcal{U}_2\circ \Gamma\circ\mathcal{U}_1[\alpha]$, that expands to $\sum_j\ketbra{\phi_j}{j}\left[ \sum_m K_m\left( \sum_i \ketbra{i}{\psi_i}\alpha\ketbra{\psi_i}{i} \right) K_m^\dagger \right]\ketbra{j}{\phi_j}$.
    Note that for higer-dimensional systems we used the Pin Map to represent the effect of the collapse of the aniclla-system wavefunction, that is $\Lambda_\text{cmplt}[\alpha]=\Tr[\alpha]\beta$.
    
    \item  Initialize as $\rho_S=\Omega[\rho_{in}]$, $\mathcal{N}_1=\Omega_1$, $\mathcal{N}_2=\Omega_2$, $(\eta, U_{SA})$, $c_1=c_2=0$\\
    For $i = 1$ to $n$:\\
    \{  \\$\rho_{SA} = U_{SA}\big(\rho_S \otimes \eta\big)U_{SA}^{\dagger}$,\\
    $\rho^1_{SA} = U_{SA}\big(\mathcal{N}_1[\rho_{in}] \otimes \eta\big)U_{SA}^{\dagger}$,\\
    $\rho^2_{SA} = U_{SA}\big(\mathcal{N}_2[\rho_{in}] \otimes \eta\big)U_{SA}^{\dagger}$,\\\\
    
    $\alpha_{SA}=(\Omega\otimes\openone )[\rho_{SA}]$, \\
    $\alpha^1_{SA}=(\Omega_1\otimes\openone )[\rho^1_{SA}]$, \\
    $\alpha^2_{SA}=(\Omega_2\otimes\openone )[\rho^2_{SA}]$, \\\\
    
    $(\Pi_1,\Pi_2)=$ \\discrim$(\text{Tr}_S[\alpha_{SA}],\text{Tr}_S[\alpha^1_{SA}],\text{Tr}_S[\alpha^2_{SA}])$,\\
    $p_1 = \text{Tr}[(\openone\otimes\Pi_1)\,\alpha_{SA}]$\\
    $p_2 = \text{Tr}[(\openone\otimes\Pi_2)\,\alpha_{SA}]$\\
    
    $count=$ rand$(p_1,p_2)$ \\\\
    
    if $count==1$\\
    then $c_1=c_1+1$, $\Pi = \Pi_1$ \\
    else $c_2=c_2+1$, $\Pi = \Pi_2$ \\\\

    $\beta_{SA}=\frac{(\openone\otimes\Pi)\,\alpha_{SA}\,(\openone\otimes\Pi)}{\text{Tr}[(\openone\otimes\Pi)\,\alpha_{SA}\,(\openone\otimes\Pi)]}$\\
    $\beta^1_{SA}=\frac{(\openone\otimes\Pi)\,\alpha^1_{SA}\,(\openone\otimes\Pi)}{\text{Tr}[(\openone\otimes\Pi)\,\alpha^1_{SA}\,(\openone\otimes\Pi)]}$\\
    $\beta^2_{SA}=\frac{(\openone\otimes\Pi)\,\alpha^2_{SA}\,(\openone\otimes\Pi)}{\text{Tr}[(\openone\otimes\Pi)\,\alpha^2_{SA}\,(\openone\otimes\Pi)]}$\\\\

    $\Lambda^1_{cmplt}=\text{map-complete}(\text{Tr}_A\,\alpha^1_{SA},\text{Tr}_A\,\beta^1_{SA})$\\
    $\Lambda^2_{cmplt}=\text{map-complete}(\text{Tr}_A\,\alpha^2_{SA},\text{Tr}_A\,\beta^2_{SA})$\\\\

    \text{Generate in vectorized form the maps:}\\
    $\Xi_1[\bullet]=\text{Tr}_A\Big[\big(\Omega_1\otimes\openone)\big[U_{SA}(\bullet\otimes\eta)U_{SA}^{\dagger}\big]\Big]$\\
    $\Xi_2[\bullet]=\text{Tr}_A\Big[\big(\Omega_2\otimes\openone)\big[U_{SA}(\bullet\otimes\eta)U_{SA}^{\dagger}\big]\Big]$\\\\
    
    $\beta_S=\text{Tr}_A\big[\beta_{SA}\big]$,\\ $\mathcal{N}_1=\Lambda^1_{complt}\circ\Xi_1\circ\mathcal{N}_1$, \\$\mathcal{N}_2=\Lambda^2_{complt}\circ\Xi_2\circ\mathcal{N}_2$\\\\

    if $c_1 > c_2$  or $10(c_1-c_2)==count-2$\\
    $\Lambda_\text{Petz}=$ Petz$(\rho_{in}, \mathcal{N}_1)$\\
    else if $c_1 < c_2$  or $10(c_1-c_2)==count-1$\\
    $\Lambda_\text{Petz}=$ Petz$(\rho_{in}, \mathcal{N}_2)$\\\\
    $(\eta,U_{SA})=$ collision$(\Lambda_\text{Petz})$ \\\\

    $\rho_S=\Lambda_\text{Petz}[\beta_S]$\\
    $\mathcal{N}_1=\Lambda_\text{Petz}\circ\mathcal{N}_1$\\
    $\mathcal{N}_2=\Lambda_\text{Petz}\circ\mathcal{N}_2$\\\\

    $F_i=\text{Tr}(\sqrt{\rho_{in}}\,\rho_S\sqrt{\rho_{in}})$\\
 
    $U_{SA}$  = SWAP
    \}

    \item Plot $F_i$ vs $i$ at each step of the loop above.
\end{itemize}

\section{Construction of the Kraus operators for general noise}
\label{general_kraus}
The first method to generate a random noisy channel is to model it as a random interaction with an ancilla $\rho_E$ to be traced out.
The environmental state is

\begin{equation}
\rho_E
=
p\,|0\rangle\langle0|
+
(1-p)\,|1\rangle\langle1|.
\end{equation}

The corresponding channel is

\begin{equation}
\mathcal{E}(\rho_S)
=
\operatorname{Tr}_E
\!\left[
U_{SE}
\left(
\rho_S\otimes\rho_E
\right)
U_{SE}^{\dagger}
\right].
\end{equation}

Substituting the environment state,

\begin{equation}
\mathcal{E}(\rho_S)
=
\operatorname{Tr}_E
\!\left[
U_{SE}
\left(
\rho_S\otimes
\bigl(
p|0\rangle\langle0|
+
(1-p)|1\rangle\langle1|
\bigr)
\right)
U_{SE}^{\dagger}
\right].
\end{equation}

The Kraus representation of the channel is

\begin{equation}
\mathcal{E}(\rho_S)
=
\sum_{\alpha,\beta=0}^{1}
K_{\alpha\beta}\,
\rho_S\,
K_{\alpha\beta}^{\dagger}.
\end{equation}

For the particular choice \(p=\frac{1}{2}\),

\begin{align}
K_{00}
&=
\frac{1}{\sqrt{2}}
\left(I_S\otimes\langle 0|_E\right)
U_{SE}
\left(I_S\otimes|0\rangle_E\right),
\\[2mm]
K_{10}
&=
\frac{1}{\sqrt{2}}
\left(I_S\otimes\langle 1|_E\right)
U_{SE}
\left(I_S\otimes|0\rangle_E\right),
\\[2mm]
K_{01}
&=
\frac{1}{\sqrt{2}}
\left(I_S\otimes\langle 0|_E\right)
U_{SE}
\left(I_S\otimes|1\rangle_E\right),
\\[2mm]
K_{11}
&=
\frac{1}{\sqrt{2}}
\left(I_S\otimes\langle 1|_E\right)
U_{SE}
\left(I_S\otimes|1\rangle_E\right).
\end{align}

\paragraph{Random Kraus Matrices}
The second method to construct a random noisy channel is to build a set of random matrices $A_i$ and then normalize them to fulfill the CPTP condition $\sum K_i^\dagger K_i = \mathbb{I}$.
This is only possible if the sum $\sum A_i^\dagger A_i = S$ is invertible, otherwise we generate new random matrices.
We ran our simulations taking the general noise Kraus operators acting on one qubit as:
\begin{align}
    &K_1=\begin{pmatrix}
        -0.110+0.012 i & 0.003-0.386i \\
        -0.121+0.291i &    0.065+0.012i
    \end{pmatrix} \\
    &K_2 = \begin{pmatrix}
        0.058+0.207i & -0.053-0.285i \\
-0.000+0.478i & 0.264-0.130i
    \end{pmatrix} \\
    &K_3 = \begin{pmatrix}
        -0.140-0.289 i & 0.063+0.487 i \\
        0.066-0.520 i & -0.059-0.202 i
    \end{pmatrix} \\
    &K_4 = \begin{pmatrix}
        0.440+0.083 i & 0.294+0.545 i \\
        0.111-0.150 i & -0.031-0.069 i 
    \end{pmatrix}
\end{align}

\section{Constructing collisional model from a CPTP map}\label{Stinespring dilation}
Let us consider a map $\Lambda[\rho]=\text{Tr} (\rho)\, \sigma$. Thus, we have 
\begin{align}
    \Lambda\big[\ket{0}\bra{0}\big]&=\Lambda\big[\ket{1}\bra{1}\big]=\sigma \\
    \Lambda\big[\ket{0}\bra{1}\big]&=\Lambda\big[\ket{1}\bra{0}\big]= 0
\end{align}
where
\begin{align}
    \sigma=p_a\begin{bmatrix}
        a_1 \\ a_2
    \end{bmatrix}\begin{bmatrix}
        a_1^* & a_2^*
    \end{bmatrix}+p_b\begin{bmatrix}
        b_1 \\ b_2
    \end{bmatrix}\begin{bmatrix}
        b_1^* & b_2^*
    \end{bmatrix}
\end{align}
with $|a_1|^2+|a_2|^2=|b_1|^2+|b_2|^2=1$ and $a_1^*b_1+a_2^*b_2=0$.
Thus, the Choi-Jamiolkowski matrix of $\Lambda$ is given by
\begin{align}
    C\big(\Lambda\big)=\begin{bmatrix}
        \sigma & 0 \\
        0 & \sigma
    \end{bmatrix}=\sum_{i=0}^3 p_i \text{vec}(K_i)\text{vec}(K_i)^{\dagger}
\end{align}
where we have $\text{vec}:\ket{i}\bra{j}=\ket{ji}$. Also, $p_0=p_2=p_a$, $p_1=p_3=p_b$, and,
\begin{align}\label{Kraus_recovery}
    K_0&=\sqrt{p_a}\begin{bmatrix}
        a_1 & 0 \\ a_2 & 0 
    \end{bmatrix},\,\, K_1=\sqrt{p_b}\begin{bmatrix}
        b_1 & 0 \\ b_2 & 0
    \end{bmatrix}, \nonumber\\
    K_2&=\sqrt{p_a}\begin{bmatrix}
        0 & a_1 \\ 0 & a_2
    \end{bmatrix},\,\, K_3=\sqrt{p_b}\begin{bmatrix}
        0 & b_1 \\ 0 & b_2
    \end{bmatrix}
\end{align}
The above are also the Kraus operators of the $\Lambda$. One can show that using appropriate $d-$dimensional ancilla $\rho_A\in\mathcal{P}_+(\mathcal{H}_d)$ and $2d\times 2d$ unitary $U_{SA}$, the map can be unraveled as
\begin{align}
    \Lambda[\rho]&=\text{Tr}_A \big(U_{SA}\,\rho\otimes\rho_{A}\,U_{SA}^{\dagger}\big)\\
    &=\sum_{i,j=0,1}\sqrt{\lambda_j}\bra{\lambda_i}U_{SA}\ket{\lambda_j}\,\rho\,\sqrt{\lambda_j}\bra{\lambda_j}U^{\dagger}_{SA}\ket{\lambda_i}\label{Kraus_formula}
\end{align}
where 
\begin{equation}
    \rho_A=\lambda_0\ket{\lambda_0}\bra{\lambda_0}+\lambda_1\ket{\lambda_1}\bra{\lambda_1}   
\end{equation}
It can be seen from above that the set $\{\sqrt{\lambda_j}\bra{\lambda_i}U_{SA}\ket{\lambda_j}:i,j=0,1\}$ also forms a set of Kraus operators. We find two such unravelings as given below:

\subsection{Case I: $d=2$}
Let us choose
\begin{equation}
    \rho_A=\begin{bmatrix}
        p_a & 0 \\
        0 & p_b
    \end{bmatrix}
\end{equation}
and
\begin{align}
    U_{SA}=\begin{bmatrix}
        a_1 & 0 & b_1 & 0 \\
        a_2 & 0 & b_2 & 0 \\
        0 & a_1 & 0 & b_1 \\
        0 & a_2 & 0 & b_2 
    \end{bmatrix}
\end{align}
One can show that using the above forms the Kraus operators in Eq.(\ref{Kraus_recovery}) can be recovered using Eq. (\ref{Kraus_formula}).\\

\subsection{Case II: $d=4$}
\subsubsection{Stinespring Dilation (Unitary Dilation)}
The Stinespring dilation is a mathematical way to ``expand'' a CPTP map in the Hilbert space of the system $\mathcal{H}_S$ to a Unitary evolution in an expanded space comprising an environment A (ancilla), $\mathcal{H}_S\otimes\mathcal{H}_A$. 
The first step is to build the Stinespring isometry:
\begin{align}
    V:\mathcal{H}_S \rightarrow \mathcal{H}_S\otimes\mathcal{H}_A\,, \\
    V = \sum_{i=1}^r K_i\otimes\ket{i}_A\,,
\end{align}
where $\{\ket{i}_A\}_{i=1}^r$ and $r$ is greater than or equal to the Kraus rank (rank of the Choi matrix).
This is called isometry because if $\sum_iK_i^\dagger K_i=\mathbb{I}_S$, then $V^\dagger V=\mathbb{I}_S$, and so
\begin{equation}
    \Lambda[\rho_S] = Tr_A\{V\rho_S V^\dagger\} \,.
\end{equation}
We can choose in this case $\rho_A=\ket{0}\bra{0}$. Note that here $\ket{0}$ is the ground state of a 4-dimensional space.
At this point we can define the diluted unitary from
\begin{equation}
    V\ket{\psi}_S = U_{SA}(\ket{\psi}_S\otimes\ket{0}_A)\,
\end{equation}
such that
\begin{align}
    \Lambda[\rho_S] =& \sum_{i=0}^rK_i\rho_SK_i^\dagger=Tr_A\{V\rho_SV^\dagger\} \\
    =&Tr_A\{U_{SA}\rho_S\otimes\ket{0}\bra{0}_AU_{SA}^\dagger\} \\
    =&Tr_A\left\{\sum_{i,j=0}^r(K_i\rho_SK_j^\dagger)\otimes\ket{i}\bra{j} \right\}=\Lambda[\rho_S] \,.
\end{align}
Each $K_i$ is a $d_S\times d_S$ matrix.
Then, $V$ is a $(4d_S)\times d_S$ matrix obtained by stacking the $K_i$ as blocks.
In this case, the unitary is given by
\begin{align}
    U_{SA}=\begin{bmatrix}
        K_0 & . & . & . \\
        K_1 & . & . & . \\
        K_2 & . & . & . \\
        K_3 & . & . & . 
    \end{bmatrix}
\end{align}
In the above only the first two columns of the $8\times 8$ matrix are given.
The other $d_S$ orthonormal columns can be found by Gram-Schmidt
The rest of the six columns can be found by choosing any orthonormal set of $6$ vectors which are also orthogonal to the first two columns. 
\subsubsection{Gram-Schmidt}

As for example, if we choose $a_1=b_1=1$ and $a_2=b_2=0$ in Eq. (\ref{Kraus_recovery}), we get
\begin{equation}
    U_{SA}=\begin{bmatrix}
        \sqrt{p_a} & 0 & 0 & -\sqrt{p_b} & 0 & 0 & 0 & 0\\
        0 & 0 & 0 & 0 & \sqrt{p_a} & 0 & 0 & -\sqrt{p_b} \\
        0 & 1 & 0 & 0 & 0 & 0 & 0 & 0\\
        0 & 0 & 1 & 0 & 0 & 0 & 0 & 0\\
        \sqrt{p_b} & 0 & 0 & \sqrt{p_a} & 0 & 0 & 0 & 0\\
        0 & 0 & 0 & 0 & \sqrt{p_b} & 0 & 0 & \sqrt{p_a}\\
        0 & 0 & 0 & 0 & 0 & 1 & 0 & 0\\
        0 & 0 & 0 & 0 & 0 & 0 & 1 & 0
    \end{bmatrix}
\end{equation}

\end{document}